\documentclass[11pt,a4paper]{article}                           

\usepackage{a4}          

\usepackage[UKenglish]{babel}
\usepackage{amsmath}
\usepackage{amssymb}
\usepackage[usenames]{color} 
\usepackage{multirow}
\usepackage{graphicx}
\usepackage{epstopdf}
 
\usepackage{ifpdf}
\ifpdf
  \usepackage[pdftex,
    pdftitle={Hypercharge Flux in IIB and F-theory: Anomalies and Gauge Coupling Unification},
    pdfauthor={Christoph Mayrhofer, Eran Palti, Timo Weigand},
    pdfsubject={F-theory and IIB compactifications},
    bookmarksopen, bookmarksnumbered, bookmarksopenlevel=2]{hyperref}
\fi
\def\hybrid{\topmargin -20pt    \oddsidemargin 0pt
        \headheight 0pt \headsep 0pt
        \textwidth 6.25in       
        \textheight 9 in       
        \marginparwidth .875in
        \parskip 5pt plus 1pt 
          \jot = 1.5ex
   }
\hybrid
\numberwithin{equation}{section}
\numberwithin{table}{section}

\newcommand{\beq}{\begin{equation}}  \newcommand{\eeq}{\end{equation}}
\newcommand{\bal}{\begin{aligned}}   \newcommand{\eal}{\end{aligned}}
\newcommand{\bea}{\begin{eqnarray}}  \newcommand{\eea}{\end{eqnarray}}

\newcommand{\nn}{\nonumber}

\newcommand{\tw}{\text{w}}

\newcommand{\f}{{\bf 5}}

\newcommand{\te}{{\bf 10}}

\newcommand{\be}{\begin{equation}}
\newcommand{\ee}{\end{equation}}

\newcommand{\half}{\frac{1}{2}}

\newcommand{\executeiffilenewer}[3]{%
 \ifnum\pdfstrcmp{\pdffilemoddate{#1}}%
 {\pdffilemoddate{#2}}>0%
 {\immediate\write18{#3}}\fi%
}

\begin{document}

\baselineskip=14pt
\parskip 5pt plus 1pt

\vspace*{-1.5cm}
\begin{flushright}    
  {\small
 CPHT-007.0213
  }
\end{flushright}

\vspace{2cm}
\begin{center}        
  {\LARGE Hypercharge Flux in IIB and F-theory: Anomalies and Gauge Coupling Unification}
\end{center}

\vspace{0.75cm}
\begin{center}        
 Christoph Mayrhofer$^{1}$, Eran Palti$^{2}$, Timo Weigand$^{1}$
\end{center}

\vspace{0.15cm}
\begin{center}        
  \emph{$^{1}$ Institut f\"ur Theoretische Physik, Ruprecht-Karls-Universit\"at, \\
             Heidelberg, Germany}
             \\[0.15cm]
  \emph{$^{2}$ Centre de Physique Theorique, Ecole Polytechnique,\\ 
               CNRS, 
               Palaiseau, France
  }
\end{center}

\vspace{2cm}


\begin{abstract}

We analyse hypercharge flux GUT breaking in F-theory/Type IIB GUT models with regards to its implications for anomaly cancellation and gauge coupling unification. 
To this aim we exploit the Type IIB limit and consider 7-brane configurations that for the first time are guaranteed to exhibit net hypercharge flux restriction to matter curves. We show that local F-theory models with anomalies of type $U(1)_Y-U(1)^2$ in the massless spectrum can be consistent only if such additional $U(1)$s are globally geometrically massive (in the sense that they arise from non-K\"ahler deformations of the Calabi-Yau four-fold). Further, in such cases of geometrically massive $U(1)$s hypercharge flux can induce new anomalies of type $U(1)_Y^2-U(1)$ in the massless spectrum, violating constraints in local models  forbidding such anomalies. In particular this implies that it is possible to construct models exhibiting a $U(1)_{PQ}$ global symmetry which have hypercharge flux doublet-triplet splitting and no further exotics.
We also show that the known hypercharge flux induced splitting of the gauge couplings in IIB models at {tree-level} can be reduced by a factor of 5 by employing a more F-theoretic twisting of $U(1)$ flux by hypercharge flux bringing it to well within MSSM 2-loop results. In the case of net restriction of hypercharge flux to matter curves this tree-level splitting becomes more involved, is tied to the vacuum expectation values of certain closed-string fields, and therefore gauge coupling unification becomes tied to the question of moduli stabilisation.

\end{abstract}

\clearpage


\newpage

\tableofcontents

\section{Introduction}

Some of the most appealing qualities of string theory realisations of Grand Unified Theories (GUTs) are that they offer new approaches to aspects of GUTs that are not available in their four-dimensional versions. Perhaps the most immediate question arising in a GUT is how the gauge group is broken to that of the Standard Model. A completely new, and string theoretic in that it is not four-dimensional, mechanism for achieving this was suggested relatively recently within the context of type IIB and F-theory GUTs which is to turn on a background flux for the hypercharge generator along the compact internal dimensions \cite{Beasley:2008kw,Donagi:2008kj}. What is most interesting about this mechanism is that it can be used for the canonical embedding of hypercharge inside $SU(5)$ thereby retaining the normalisation of the generator that naturally leads to gauge coupling unification.\footnote{The use of hypercharge flux to break the GUT group but without retaining the appropriate normalisation naturally is a much older idea which was suggested already in \cite{Witten:1985bz} and first used in \cite{Antoniadis:2000ena}. See \cite{Tatar:2006dc,Blumenhagen:2006ux} for a detailed discussion in the heterotic context.} The important point which allows for this possibility while keeping the hypercharge generator massless is that the GUT gauge group is localised on a submanifold of the full extra dimensions, the so-called GUT brane which we denote by $S$. This implies that there are non-trivial flux configurations supported on $S$ which do not induce a St{\"u}ckelberg mass for the hypercharge gauge potential. The conditions on the hypercharge flux such that a mass is not induced for the hypercharge gauge field were stated in \cite{Buican:2006sn} within a type IIB string theory setting. These were generalised to an F-theory setting in \cite{Beasley:2008kw,Donagi:2008kj}. 

Apart from breaking the GUT group the hypercharge flux potentially offers a solution to yet another puzzle: the absence of triplet partners to the SM Higgs doublets, known as doublet-triplet splitting \cite{Beasley:2008kw,Donagi:2008kj}. In F-theory models the Higgs fields typically localise on so-called matter curves in $S$. A non-trivial restriction of the hypercharge flux to the matter curves supporting the Higgs doublets can induce a massless spectrum which is in incomplete GUT multiplets. Studies of hypercharge flux within an F-theory context however have so far been rather implicit in the sense that they utilise a local effective gauge theory description of the F-theory model, see \cite{Weigand:2010wm,Leontaris:2012mh,Maharana:2012tu} for reviews. 
Even though certain aspects of hypercharge flux can be understood completely within a local effective theory, others really require a full global understanding of the appropriate F-theory construction. This is particularly the case for the restriction of the flux to the matter curves. And while the conditions on the hypercharge flux for $U(1)_Y$ to remain massless have been studied in the better understood type IIB context, the restriction to the matter curves has not. Indeed before this work there were no global complete models where a non-trivial net restriction of the hypercharge flux to matter curves was shown. Strictly speaking it was not even clear if consistent compactifications with this property exist, both in IIB and in F-theory. Given the difficulty of a full study in F-theory it is therefore natural to perform such a study in the type IIB setting first, and this is the primary aim of this note. By moving to a simpler, but better understood, setting we will be able to sharpen questions regarding the possible spectrum that can be induced by hypercharge flux, in particular in the presence of additional $U(1)$ symmetries. We will show that indeed hypercharge flux \emph{can} restrict non-trivially to matter curves in a \emph{consistent} setting. In turn answering these questions in detail will have important implications for model building in F-theory.

An important result of this note is a clarification of an issue raised in \cite{Palti:2012dd} regarding anomaly cancellation in the presence of hypercharge flux. The specific details are important to fully understand the question raised, how it is resolved in IIB, how this resolution is uplifted to F-theory, and finally what the implications are for model building. However before delving into the details we present the main points here. The primary issue is that in local F-theory models with $U(1)$ symmetries -  the precise nature of these symmetries will be crucial as we will see below - it was pointed out in \cite{Palti:2012dd} that the Abelian anomaly involving the hypercharge and two $U(1)$ symmetries does not automatically vanish, ${\cal A}_{U(1)_Y-U(1)^2}\neq 0$. This is surprising in the following sense: the hypercharge flux is expected not to couple to the closed-string sector in order to not induce a St{\"u}ckelberg mass for the $U(1)_Y$ gauge field. This is sometimes stated as the constraint that it should be globally trivial. However, as a consequence it cannot modify the Green-Schwarz anomaly cancellation mechanism, and therefore should not induce any new anomalies. What leads to a puzzle is that the analogous anomaly before the hypercharge flux is introduced vanishes, ${\cal A}_{SU(5)-U(1)^2} = 0$, and so it seems a new anomaly is introduced by the flux.\footnote{One might consider what happens if such a `local' $U(1)$ symmetry is spontaneously broken in the bulk only, by the vevs  of some $U(1)$ charged singlets whose matter curve has no Yukawa point intersection with the GUT divisor. Because pure bulk recombination in this sense cannot induce a mass for charged states running in the anomaly, it cannot cancel a non-vanishing anomaly of the type we are considering. If the matter curve of the recombination fields intersects the GUT brane, then charged states can become massive through Yukawa type couplings to GUT singlets. However in this case also the matter curves ceases to be split or are connected through monodromies. Put differently, we define a local $U(1)$ as one which is only (potentially) broken in the bulk.} As a result, it was conjectured in \cite{Palti:2012dd} that in globally consistent F-theory models there are additional geometric constraints which guarantee that ${\cal A}_{U(1)_Y-U(1)^2} = 0$ in the matter spectrum. Imposing this condition on the spectrum is very restrictive for model building. 

In this work we will show that the described puzzle can be recreated in type IIB string theory where the questions can be posed much more sharply because the $U(1)$ symmetries are well understood. We will find that there are models where indeed ${\cal A}_{U(1)_Y-U(1)^2}\neq 0$ and the anomaly {\it is} cancelled by the Green-Schwarz mechanism. We will show that this is possible in IIB because the hypercharge flux {\it can} couple to the closed-string sector without inducing a mass for the $U(1)_Y$ gauge potential as long as it couples only to the orientifold odd sector. Put differently, in IIB the hypercharge flux is not globally trivial. The Green-Schwarz mechanism can then cancel the anomaly as long as the additional $U(1)$s are what was called geometrically massive in \cite{Grimm:2010ez,Grimm:2011tb}, which means that they have a St{\"u}ckelberg mass even in the absence of any flux background. The uplift of such geometrically massive $U(1)$s and the associated fluxes to F-theory is not well understood but some work on it has been initiated in \cite{Grimm:2010ez,Grimm:2011tb,Krause:2012yh,Grimm:2013fua}. In particular it has been conjectured that geometrically massive $U(1)$s are associated to non-K{\"a}hler deformations of the M-theory dual. The conclusion we draw for F-theory model building is that the constraint ${\cal A}_{U(1)_Y-U(1)^2} = 0$ in the matter sector can be relaxed, at least partially if not fully, in the case where the $U(1)$s are precisely of this type. In all other cases we expect that an explicit analysis of the interplay of matter curves and hypercharge flux in a full global setup will ensure absence of the anomalies.

Analogously we will show further that results from local models found in \cite{Dudas:2010zb} which were interpreted in \cite{Marsano:2010sq} as constraints which ensure that the following anomalies are proportional 
\be
{\cal A}_{SU(3)^2-U(1)} \propto {\cal A}_{SU(2)^2-U(1)} \propto {\cal A}_{U(1)_Y^2-U(1)} \propto {\cal A}_{{\bf SU(5)^2}-U(1)} \;,
\ee
can be relaxed in the cases where the $U(1)$s are geometrically massive. This has important implications for model building and in particular potentially allows to avoid the problem of exotics in the presence of a $U(1)_{PQ}$ raised in \cite{Dudas:2010zb}. Indeed we will present a toy model which has a (massive) $U(1)_{PQ}$ symmetry, doublet-triplet splitting by hypercharge flux, and no additional exotics.

In \cite{Blumenhagen:2008aw} another crucial aspect of hypercharge flux was found which is that it splits the gauge couplings at tree level through $F^4$ type terms in the D7 DBI action. The calculation performed was in a type IIB setting though it is expected that such a splitting survives in F-theory up to possible modifications by $D(-1)$ instantons and further threshold effects. Within the IIB setting it was shown that hypercharge flux splits the gauge couplings at roughly the 4-5\% level, which is slightly too large to be consistent with MSSM results of a 3\% splitting. As one of the results of this paper we will show that this splitting crucially relies on a particular choice of twisting by the hypercharge flux. This choice is not the only possibility and there exists an alternative choice which is the one more naturally associated to F-theory. This latter choice we find reduces the splitting by a factor of 5 leading to an estimate of 1\% splitting, much smaller than MSSM 2-loop results. Using our understanding of the restriction of hypercharge flux to matter curves we will show that in the case of non-trivial restriction the splitting is modified further by $F^3$ terms which can induce an arbitrarily large splitting, the precise value depending on the full moduli stabilisation prescription. Fortunately this splitting can vanish within appropriate and natural frameworks. These results are discussed in section \ref{sec:hypgauge}.

We now go on to review in more detail the properties of hypercharge flux in F-theory before studying in the next section hypercharge flux in type IIB string theory.

\subsection{Hypercharge flux in F-theory -  Local picture and anomalies}
\label{sec:hypf}

Since we are primarily motivated by F-theory constructions we begin by summarising some of the relevant results in this setting. This will set the background for the type IIB study that we perform. We refer to \cite{Weigand:2010wm,Maharana:2012tu} for reviews of the following. The constraints on the hypercharge flux for the $U(1)_Y$ gauge potential to remain massless in F-theory were presented in \cite{Beasley:2008kw,Donagi:2008kj}. 
In \cite{Beasley:2008kw}, and most of the ensuing local model building literature,  the hypercharge flux in F-theory models is specified simply by a line bundle in the gauge theory on $S$, as in IIB. A full F-theoretic description should specify instead a four-form $G_4$-flux defined as an element of $H^{2,2}(\hat Y_4)$, where $\hat Y_4$ is the resolved Calabi-Yau four-fold. We will provide such a description in section \ref{sec_hyperG4}, which will turn out  to be very similar to the analysis in \cite{Donagi:2008kj} of hypercharge $G_4$ in the language of a \emph{local} ALE fibration over the GUT divisor $S$. 

In any case, either by analogy with IIB or via an explicit derivation, hypercharge flux can be described by an element $f_Y \in H^2(S)$.
The condition for masslessness is that the flux should have vanishing intersection with the pullback of any globally non-trivial element of the cohomology class $H^2(B)$ on base $B$ of the elliptic fibration  - see \cite{Beasley:2008kw,Donagi:2008kj} or section \ref{sec_hyperG4} for derivation in the respective approaches. This condition is sometimes referred to as global triviality of the hypercharge flux.

 It is worth being precise about what this means. Consider the embedding $\iota: S \rightarrow B$ of the GUT surface $S$ into the three-fold base $B$. This induces a map 
\bea
\iota^*: H^2(B) \rightarrow H^2(S) \;
\eea
for the pullback of cohomology from $B$ to $S$
as well as the pushforward on homology as
\bea
\iota_*: H_2(S) \rightarrow H_2(B) .
\eea
We can think of the latter as the embedding of curves in $S$ into $B$. This allows us to introduce the Gysin map, i.e.\ the pushforward on cohomology, as the map
\bea
\iota_!: H^2(S) \rightarrow H^{4}(B) \;
\eea
defined by taking the Poincar\'e dual on $S$, applying $\iota_*$ and dualising again on $B$. The pushforward and pullback on cohomology satisfy the relation
\bea
\int_B \iota_! \Omega \wedge \omega = \int_S \Omega \wedge \iota^*\omega \qquad \forall \;\omega \in H^{2}(B)\;,\;\; \Omega \in H^2(S) \;.
\eea
With this expression we see that the condition for the hypercharge flux to wedge to zero with the pullback of any globally non-trivial two-form implies that its pushforward $i_! F_Y$ must wedge to zero with all elements in $H^2(B)$ and therefore should be in the trivial class.

Now we turn to the matter curves in F-theory. The elliptic fibration of an $SU(5)$ F-theory GUT model can be brought globally, at least to leading order in $w$ \cite{Katz:2011qp}, to the Tate form
\be
-y^2 + x^3 + b_5 xyz + b_4 w x^2 z^2 + b_3 w^2 y z^3 + b_2 w^3 x z^4 + b_0 w^5 z^6 = 0 \;.
\ee
The elliptic fibre is described by this equation in the space $\mathbb{P}_{[2,3,1]}$ with homogeneous coordinates $\left\{x,y,z\right\}$. The coefficients $w$ and $b_i$ are sections of certain bundles in the base $B$ of the fibration, pulled back to the four-fold $Y_4$. Here $w=0$ specifies the GUT divisor whose projection to $B$ gives the 4-cycle $S$. There is an $SU(5)$ singularity at $x=y=0$ in the fiber over $w=0$, and this singularity enhances over loci where additionally some combinations of the $b_i$ vanish. Since this imposes one additional equation in the base these loci are curves on $S$. As the enhancement of the singularity implies that additional massless matter is localised there, they are denoted matter curves. The simplest such curves are where $b_5=0$ and the singularity enhances to $SO(10)$ and therefore states in the the $\te$ representation of $SU(5)$ localise there: ${\cal C}_{\te} = S \cap \left\{b_5=0\right\}$.

The net restriction of the hypercharge flux to the matter curves, which determines the net non-GUT multiplets on them, is given by the expression
\be
\int_S f_Y \wedge \iota^*\left[b_5\right] \;.
\ee
Here $\iota^*$ denotes the pullback to $S$ of the two-form $\left[b_5\right]$ which is the Poincar\'e dual on $B$ of the divisor $b_5=0$, and we denote the hypercharge flux by $f_Y$. This is the guess one would make from the gauge theory description. We can now directly obtain the first simple result regarding hypercharge restriction to matter curves which is that in this model it can have no net restriction to the matter curve since $\left[b_5\right]$ is a globally non-trivial class in $B$ \cite{Donagi:2008kj,Andreas:2009uf}. It is simple to see that the same conclusion holds for the $\f$-matter curves as well where $\left[b_5\right]$ is replaced by $\left[b_3^3 b_4 - b_2b_3b_5 +b_0b_5^2 \right]$, which is the locus where the gauge group enhances to $SU(6)$.

With this in mind it is not immediate to see how the hypercharge flux can even in principle have net restriction to the matter curves. A way out of this was proposed in \cite{Donagi:2009ra,Marsano:2009gv}, in a local framework. The point is that in some cases the restriction of $b_5$ to $S$ can be such that it splits into a number of components
\be
\left.b_5\right|_{w=0}=\prod_i c_i \;,\;\; c_i \in H^2(S) \;. \label{factorb5}
\ee
Now it is only the product of the $c_i$, or the homological sum if we think in classes, which must come from the pullback of a non-trivial element of the bulk. The Poincar\'e dual two-form of each of the $c_i$ on $S$ can have a component which not in the image of $i^*$ acting on elements of $H^2(B)$, but the sum over such components of all the $c_i$ must vanish. Now the hypercharge flux can have \emph{in principle} non-trivial intersection with the $[c_i]$
\be
\int_S f_Y \wedge \left[c_i\right] \neq 0\;.
\ee
The interpretation of such a possible split in terms of the gauge theory is that there is a $U(1)$ symmetry under which the matter localised on the different $c_i$ factors have different charges. Such constructions are often terms split spectral cover models and have been studied extensively in the literature, see  \cite{Maharana:2012tu} for a review. Note  that since the splitting occurs upon restricting to $S$, the associated $U(1)$ symmetry is a local one in the sense that whether it is preserved in the full global compact model remains unknown. 
While (\ref{factorb5}) is in principle compatible with a non-trivial hypercharge flux restriction to the curves, more work is required to determine if the this restriction is actually non-vanishing.
In particular in such a local framework a complete analysis of the \emph{consistent} types of splittings cannot be provided. As discussed below this is demonstrated by the appearance of extra field theoretic consistency conditions which had not been detected by the local geometric analysis.

A general procedure to globally complete such local $U(1)$s was proposed in \cite{Mayrhofer:2012zy},  generalizing previous constructions of massless $U(1)$ symmetries in F-theory GUTs \cite{Grimm:2010ez,Krause:2011xj,Grimm:2011fx}. The idea is to impose the factorisation of type (\ref{factorb5}) to hold to all globally, which means to all orders in $w$. It was shown then that the local $U(1)$s give rise to global massless $U(1)$s. E.g. the  $SU(5) \times U(1)$ models of \cite{Mayrhofer:2012zy} are realized as $\mathbb P_{1,1,2}[4]$-fibrations (see also \cite{Morrison:2012ei}). A different class of $SU(5) \times U(1)$ models, given by a $\mathbb P_{1,1,1}[3]$-fibration, was found in \cite{Braun:2013yti}.
As it stands, all of these existing models share the important property that without further modifications the components of the matter curves $c_i$ all come from globally non-trivial classes and therefore can have no net hypercharge restriction due to splitting as in (\ref{factorb5}).

Returning to the general models with only local splitting, the fact that the hypercharge flux must have no net restriction to the pull backs of the $[b_i]$ implies certain constraints on its possible net restriction to the $[c_i]$ and therefore to the matter curves. These were studied in \cite{Marsano:2009gv,Marsano:2009wr,Dudas:2009hu,Palti:2012aa} and in particular 
in \cite{Dudas:2010zb} it was shown for a general class of models (and later proven for all models in \cite{Dolan:2011iu}) that the following constraints hold on the restriction to the matter curves
\be
\sum_i Q^i_A \int_S f_Y \wedge \left[C_{\f^{(i)}}\right] + \sum_j Q^j_A \int_S f_Y \wedge \left[C_{\te^{(j)}}\right] = 0 \;. \label{dp}
\ee
Here we have introduced the notation for the matter curves $C_{\f^{(i)}}$ and $C_{\te^{(j)}}$ which for $SU(5)$ models carry matter in the $\f$ and $\te$ representations respectively. The expressions $\left[C_{\f^{(i)}}\right]$ denote the two-forms Poincar\'e dual on $S$ to the curves. The charges $Q_A^i$ are those of the matter localised on the curves under any combination of the local $U(1)$s that correspond to the splitting structure (\ref{factorb5}).
Similarly we have the constraints just discussed that the sum of the net restriction to the matter curves must vanish
\be
\sum_i \int_S f_Y \wedge \left[C_{\f^{(i)}}\right]  = \sum_j \int_S f_Y \wedge  \left[C_{\te^{(j)}}\right] = 0 \;. \label{nosum}
\ee

In \cite{Marsano:2010sq} the geometric constraints (\ref{dp}) and (\ref{nosum}) were given a physical interpretation in terms of four-dimensional anomaly cancellation. The crucial point raised is that the global triviality of the hypercharge flux implies that it cannot modify the Green-Schwarz mechanism and therefore the anomalies in the massless spectrum after turning on hypercharge flux must be proportional to the anomalies without hypercharge flux. The constraints on the spectrum coming from (\ref{dp}) and (\ref{nosum}) precisely ensure this for some of the anomalies. In \cite{Palti:2012dd} it was shown that (\ref{dp}) and (\ref{nosum}) however do not ensure that all the anomalies are proportional to the GUT ones, in particular the anomaly ${\cal A}_{U(1)_Y-U(1)^2}$, and therefore a puzzle arose as to what guarantees full anomaly cancellation. Concretely the anomalies map is
\bea
(\ref{dp}) &\implies & {\cal A}_{SU(3)^2-U(1)} \propto {\cal A}_{SU(2)^2-U(1)} \propto {\cal A}_{U(1)_Y^2-U(1)} \propto {\cal A}_{{\bf SU(5)^2}-U(1)} \;,  \label{YY1a}\\
(\ref{nosum}) &\implies & {\cal A}_{SU(3)^2-SU(3)} \propto ... \propto {\cal A}_{U(1)_Y^2-U(1)_Y} \propto {\cal A}_{{\bf SU(5)^2-SU(5)}}=0 \;, \\
 (???) &\implies & {\cal A}_{U(1)_Y-U(1)^2} \propto {\cal A}_{SU(3)-U(1)^2} \propto {\cal A}_{SU(2)-U(1)^2} \propto {\cal A}_{{\bf SU(5)
}-U(1)^2} = 0 \;. \label{puzan}
\eea
Following the logic that the hypercharge flux should not modify anomaly cancellation it was therefore proposed in \cite{Palti:2012dd} that the hypercharge is restricted so as to ensure that ${\cal A}_{U(1)_Y-U(1)^2}=0$. This amounts to the constraint
\be
\sum_i Q_i^A Q_i^B \int_S f_Y \wedge \left[C_i^{\f}\right] + 3 \sum_j Q_j^A Q_j^B \int_S f_Y \wedge \left[C_j^{\te}\right] = 0 \;, \label{p}
\ee
where $Q_j^A$ and $Q_j^B$ denote the charges under any 2 combinations of local $U(1)$s. This additional constraint was shown to be very restrictive on the possible models that could support net hypercharge flux restriction. In particular it raises the question about the geometric consistency conditions governing the splitting of curves and the restriction of hypercharge flux.  

The primary motivation for this work is to study this issue in a type IIB framework where we have much better control and understanding of the $U(1)$ symmetries. We will show that indeed for some models the relation (\ref{puzan}) is not satisfied. We will then explain how this puzzle is resolved in the type IIB framework and therefore shed light on the uplift of this solution to F-theory. We will see that imposing (\ref{p}) on the spectrum is not always necessary thereby relaxing the stringent constraints on model building in F-theory, and further state exactly when this constraint can be relaxed. As a further important result we will show that (\ref{YY1a}) can also be violated thereby implying that in some models (\ref{dp}) does not hold. It is important to note that the hypercharge restriction we will find in IIB is not nescessarily related to the splitting (\ref{factorb5}). Indeed it is more likely that a whole new mechanism of hypercharge restriction in F-theory is avaialble as the uplift of our IIB results.

\section{Hypercharge flux in type IIB string theory}

In this section we study hypercharge flux in type IIB string theory. Although this is a significantly simpler setting than within F-theory models it has the crucial advantage that $U(1)$ symmetries are very well understood. In turn this means that the definition of the matter curves from a global perspective is sharper than in F-theory. Indeed in thinking about the restriction of the hypercharge flux to matter curves in IIB compared with F-theory one is almost immediately faced with a puzzle: since the $U(1)$ branes in IIB are localised on globally non-trivial divisors the matter curve classes are by definition pullbacks of globally non-trivial bulk forms. How can we have then any net hypercharge restriction to them at all? The answer to this is very clean in IIB and is discussed in section \ref{sec:iibhypmat}. In section \ref{sec:iibhypano} we study the implications for anomaly cancellation and exemplify, in section \ref{sec_311}, our findings in a family of brane setups with $SU(5) \times U(1)$ gauge symmetry which generalise possible charge assignments that have appeared in the F-theory literature. But first we summarise the known constraints on hypercharge masslessness emphasising the important details.

\subsection{Masslessness constraints}

We consider $SU(5)$ GUT models in the context of Type IIB orientifold compactifications on a Calabi-Yau three-fold $X$. The  gauge group $U(5) = SU(5) \times U(1)_S$  arises from a stack of five 7-branes along a holomorphic divisor $S \subset X$ and its orientifold image stack on $S'$. In addition the compactification contains extra branes along divisors $D_i$ and $D_{i'}$ such as to satisfy the D7-brane tadpole cancellation condition 
\bea \label{D7tad}
 5(S + S') + \sum_i (D_i + D_{i'}) = 8 O7
\eea
in homology. Explicit constructions of this type of Calabi-Yau orientifolds have been worked out in detail in \cite{Blumenhagen:2008zz}, to which we refer for more details.

Our notation does not distinguish between a divisor and its Poincar\'e dual two-form. It will prove convenient to introduce the index $I = 0, i$ such that $D_0 = S$. For simplicity we do not consider additional non-Abelian gauge groups so that all extra brane stacks consist of single branes. 
The described setup allows for a rich number of realisations corresponding to different numbers of extra $U(1)$ branes. Note that one or more of the brane pairs along $D_i + D_{i'}$ can be replaced by  7-branes along invariant divisors $W_i$ of Whitney type, which carry no $U(1)$ gauge group. In a \emph{generic} model (\ref{D7tad}) is the only homological relation between the brane divisors. 

The idea of hypercharge flux breaking is to embed a line bundle $L_Y$ on $S$ into $SU(5)$ by identifying the generator of its structure group with the hypercharge generator $T_Y$ of $SU(5)$. The corresponding hypercharge flux is described by the first Chern class  $c_1(L_Y) \equiv f_Y  \in H^2(S) $ and breaks $SU(5) \rightarrow SU(3) \times SU(2) \times U(1)_Y$.
From the GUT brane stack $S$ therefore two $U(1)$ field strengths arise and we decompose
\bea \label{norm4D}
F_S = F_S \, T_S^{0} + F_Y \, T_Y \equiv  F_S \, \mathrm{diag}\left(1,1,1,1,1\right) +  \frac16 F_Y \, \mathrm{diag}\left(-2,-2,-2,3,3\right)\;,
\eea
where $F_S$ and $T_S^{0}$  refer to the diagonal $U(1)_S$ of $U(5)$. 
In addition the Abelian sector contains the diagonal $U(1)_i$ from the extra branes along $D_i + D_{i'}$ with associated field strengths $F_{i}$ and generators $T_i^{0}$ so that
\bea
{\rm tr }\left(T_Y\right) = 0, \qquad {\rm tr}\left(T_S^{0}\right)= 5, \qquad {\rm tr }\left(T_i^{0}\right) = 1.
\eea
For later use we note that with the above normalisation of the four-dimensional $U(1)$ field strengths the $SU(5)$ representations decompose into  $SU(3) \times SU(2) \times U(1)_Y$ representations as
\bea \label{decomp-rep}
&& {\bf 5} \rightarrow {\bf 3}_{-{1}/{3}} + {\bf 2}_{{1}/{2}}, \qquad {\bf 10} \rightarrow ({\bf 3}, {\bf 2})_{{1}/{6}} + ({\bf \bar 3},1)_{-{2}/{3}} + (1,1)_1, \\
&& {\bf 24} \rightarrow {\bf 8}_0 + {\bf 3}_0 + {\bf 1}_0 + ({\bf 3}, {\bf 2})_{-5/6} + ({\bf \bar 3}, {\bf 2})_{5/6}.
\eea
To discuss the Abelian gauge symmetries we need to recall briefly the most important features of the St\"uckelberg mechanism in compactifications with 7-branes.
The St\"uckelberg masses for the Abelian gauge factors are derived by dimensional reduction of the 7-brane Chern-Simons (CS) action
\bea \label{SCS1}
S_{CS} &=&\mu_7   \int_{D7} \Big(  {\rm tr}\, e^{F} \sum_{p}   \iota^*C_{2p}   \Big)  \sqrt{\frac{\hat A(TD7)}{\hat A(ND7)}},
\eea
where the last factor denotes the curvature terms, which will not be of interest to us.
For a detailed exposition of this derivation and a list of original references on the mixed Green-Schwarz mechanism see \cite{Plauschinn:2008yd,Grimm:2011tb}. 
The dimensional reduction expands the RR-fields in terms of a basis  $\omega_\alpha, \alpha = 1, \ldots h^{1,1}_+(X)$ and $\omega_a, a = 1, \ldots h^{1,1}_-(X)$ of the orientifold even and odd cohomology groups $H^{1,1}_\pm(X)$ as well as the dual four-forms  $\tilde \omega^\alpha$ of $H^4_+(X)$ and $\tilde \omega^a$ of $H^4_-(X)$ as
\bea \label{expansion1}
C_6 &=&  c^2_a \,   \wedge \,  \tilde \omega^a + \ldots, \qquad C_4 = c^\alpha_2\,  \wedge \,  \omega_\alpha + c^0_\alpha \, \tilde \omega^\alpha + \ldots, \\
C_2 &=& c_0^a \,  \omega_a + \dots .
\eea
By slight abuse we denote the 4-dimensional field strengths by the symbols $F_I, F_S, F_Y$ and the corresponding internal fluxes by $f_I, f_S, f_Y$.
The St\"uckelberg mass terms result from 4-dimensional couplings of the type $F  \wedge (\rm two-form)$. Two very different contributions $S^1_{St}$ and $S^2_{St}$ of  this type follow from the CS-action, namely
\bea
S^1_{St} &\simeq& \sum_I {\rm tr} \left(T_I^{0}\right) C^a_I \int_{\mathbb R^{1,3}}  F_I \wedge c^2_a, \label{St1} \\
S^2_{St} &\simeq&   \int_{\mathbb R^{1,3}} {\rm tr} (T_Y^2) F_Y \wedge c_2^\alpha \int_{S} f_Y \wedge \iota^*\omega_\alpha + \nonumber \\
 && \sum_{I} \int_{\mathbb R^{1,3}} {\rm tr} (T_I^{0})^2 F_I \wedge c_2^\alpha \int_{D_I} f_I \wedge \iota^*\omega_\alpha \label{St2},
\eea
where the Poincar\'e dual two-form to the divisor $D_I$ is $D_I= C_I^a \omega_a + C_I^\alpha \omega_\alpha$.

Note that  (\ref{St1}) is independent of any fluxes and thus gives rise to a geometric mass term in the nomenclature of \cite{Grimm:2011tb}, which also initiated the study of analogous effects in F-theory.
Since ${\rm tr} \, T_Y = 0$, hypercharge cannot acquire a geometric St\"uckelberg mass term (\ref{St1}).  In order for $U(1)_Y$ to remain entirely massless one must also avoid the flux-induced St\"uckelberg mass (\ref{St2}). This amounts to requiring that 
\bea \label{orthY}
\int_S f_Y \wedge \iota^*\omega_\alpha = 0 \qquad \forall \quad \omega_\alpha \in H^2_+(X).
\eea
The fact that $f_Y \in H^{2}(S)$ must have vanishing intersection with $\iota^* H^2_+(X)$  does not preclude a non-zero intersection with the pullback of orientifold-odd two-forms $\omega_a \in H^2_-(X)$. In particular, in Type IIB orientifolds, unlike in F-theory, hypercharge flux is by no means ``globally trivial'', by which one usually means that $f_Y$ would lie in the kernel of the pushforward map $\iota_!: H^2(S) \rightarrow H^4(X)$. This important fact was noted already in \cite{Buican:2006sn}, but in this paper we will present the first analysis of its implications for the restriction to matter curves.

\subsection{Restriction to matter curves and chirality}
\label{sec:iibhypmat}

In type IIB the matter curves are the intersection of bulk $U(1)$ branes with the $SU(5)$ stack, and therefore their classes are by definition pullbacks of cohomologically non-trivial bulk two-forms.
 A hypercharge flux satisfying (\ref{orthY}) can in principle induce non-trivial chiralities, but only along matter curves on $S$ whose dual classes $C \in H^2(S)$ do not lie completely in the image under pullback $\iota^*$ of $H^2_+(X)$. 

The matter curves are of the following form:
The states in the ${\bf 10}$-representation of $SU(5)$ arise exclusively at the intersection of the divisor $S$ with $S'$. Note that since the components of $S \cap S'$ away from the O7-plane give rise to states both in the ${\bf 10}$- and the ${\bf 15}$-representation, we assume that no such components of $S \cap S'$ exist  to avoid the exotic ${\bf 15}$-states.  Therefore the ${\bf 10}$-curve is $C_{\bf 10} = S \cap S'= S \cap O7$ and thus in perturbative models without ${\bf 15}$-representation
\bea \label{10chir}
\int_{C_{\bf 10}} f_Y = \int_S f_Y \wedge \iota^* O7 =  0.
\eea
The last equality is a consequence of (\ref{orthY}) because the class of the O7-plane is in $H^{1,1}_+(X)$. 
For later purposes we note that absence of ${\bf 15}$-states also implies that
\bea
\int_S f_Y \wedge \iota^*S = \int_S f_Y \wedge \iota^*(S + S') - \int_S f_Y \wedge \iota^*S' = \int_S f_Y \wedge \iota^*(S + S') - \int_S f_Y \wedge \iota^*O7 =0. \nonumber \\ \label{YSS0}
\eea

States in the ${\bf 5}$-representation localise on
\bea \label{SDi}
&& S \cap D_i = \frac{1}{2} (S \cap D_i^+ + S \cap D_i^-), \qquad D^\pm = D \pm D', \\
&&   S \cap D_{i'} = \frac{1}{2} (S \cap D_i^+ - S \cap D_i^-).
\eea
The dual two-form classes as elements in $H^2(S)$ have components in the pullback both of $H^{1,1}_+(X)$ and of $H^{1,1}_-(X)$. 
Only the latter components lead to non-trivial chirality in the presence of massless hypercharge flux. We exemplify this in section \ref{sec_311}.

The perturbative IIB $SU(5)$ GUT models constructed in this way differ in two aspects from their more general F-theory counterparts: First the states in the ${\bf 10}$-representation all carry the same  charge  under extra Abelian gauge groups in the model because they arise exclusively at the intersection of the divisor $S\cap O7$. In F-theory GUT models, on the other hand, several ${\bf 10}$-curves with different $U(1)$ charges are possible. The first examples of such F-theory compactifications with several ${\bf 10}$-curves have been presented recently in \cite{Mayrhofer:2012zy}.
Second, the ${\bf 10 \, 10 \, 5}$ Yukawa coupling is absent perturbatively, while in generic F-theory models this type of couplings is associated with $E_6$-enhancements over points in the base $B$. 
Thus the Type IIB models under consideration correspond to a special subset of F-theory models with a single type of ${\bf 10}$-matter curves and where the $E_6$-enhancement points are absent as a consequence of the intersection numbers of certain divisor classes \cite{Krause:2012yh}. 
The presence of the Yukawas is most likely not of relevance for the questions of interest in this note.

\subsection{Hypercharge anomalies}
\label{sec:iibhypano}

We now analyze possible hypercharge anomalies in a consistent 7-brane setup.
We are interested in the mixed anomalies of a linear combination
\bea
U(1)_A = \sum_i Q^i_A U(1)_i + Q^S_A U(1)_S =  \sum_I Q^I_A U(1)_I
\eea
of the diagonal Abelian groups $U(1)_i$ and $U(1)_S$ with $U(1)_Y$.
Such a $U(1)_A$  is massless if $Q_A^I$  lies in the kernel of both matrices
\bea \label{Massmatrix}
M_{I \alpha} = {\rm tr}\left(T_I^{0}\right)^2 \int_{D_I} F_I \wedge \iota^* \omega_\alpha,       \qquad M_{I}^a =  {\rm tr} \left(T_I^{0}\right) C_I^a.
\eea
 In particular, in the absence of any gauge flux, a linear combination $U(1)_A$ of the diagonal $U(1)$ groups is massless if 
 \bea
 \sum_I {\rm tr} \left( T^{(0)}_I \right)C_I^a Q^I_A =0.
 \eea
It is these linear combinations which correspond to massless (in the absence of gauge flux) $U(1)$s also in F-theory.

Conversely, given the natural splitting of the $U(1)$s into those with no components along $S$ and $U(1)_S$ it is worth noting that in a \emph{generic} IIB compactification with an $SU(5)$ GUT and no other non-Abelian symmetries all the $U(1)_{A}$ with $Q_A^S=0$ are geometrically massive. By generic we mean that the only homological relation satisfied by the brane divisors is the one implied by 7-brane tadpole cancellation (\ref{D7tad}).
Let us define a single divisor class $D_A$ as a suitable linear combination $Q_A^i D_i$ of the brane divisor classes and associate to it the Abelian group $U(1)_A$. Then one can solve (\ref{D7tad}) for one of the $D_k$ with $Q_A^k \neq 0$ in homology and finds 
\be
D_A = Q_A^k (4 O7 + a S + b S'  + ...) \;,
\ee
and analolgously for $D_{A'}$. Since $a+b=5$ we see that $a-b \neq 0$ and so the odd component
\be
D_A - D_{A'}  \simeq  \left(a-b\right)\left(S-S' \right) + ...
\ee
always has a component along $S-S'$, which is a homologically non-trivial class. It therefore couples to the appropriate axion in the geometric mass term. We stress again that this assumes a generic configuration in the sense that no other divisor classes have further homological relations involving $S$ which could cancel this dependence.
 
As is well-known, the generalized Green-Schwarz mechanism combines the 4-dimensional St\"uckelberg couplings involving one of the two-forms $c^2_a$ or $c_2^\alpha$ with couplings of the dual axion to two gauge or curvature field strengths, schematically 
\bea
( F_I \wedge c_2^\alpha)  -  (c^0_\alpha   \, F_J \wedge F_K) \qquad {\rm or} \qquad (F_I \wedge c^2_a) -  (c_0^a   \, F_J \wedge F_K).
\eea
This gives rise to cubic terms which cancel the corresponding 4-dimensional mixed cubic anomalies. 
For a detailed account of the resulting Green-Schwarz (GS) terms and the cancellation of mixed anomalies associated with the diagonal $U(1)_I$ we refer to \cite{Plauschinn:2008yd}. This reference considers anomaly cancellation for Abelian gauge groups in the presence of diagonally embedded flux $f_I$. For this setup, \cite{Plauschinn:2008yd} demonstrates explicitly that every anomalous diagonal $U(1)_I$ is necessarily St\"uckelberg massive and that the Green-Schwarz terms cancel the anomaly provided D7-brane and D5-tadpole cancellation conditions are satisfied.

What we are interested in are the subtle consequences of the Green-Schwarz mechanism for mixed anomalies involving $U(1)$ gauge groups that arise for hypercharge flux in $SU(5)$ GUT models. 
As a warm-up let us first demonstrate the cancellation of the anomaly ${\cal A}_{U(1)_Y^2-U(1)}$ of (\ref{YY1a}), which is well understood also in F-theory.
Since $U(1)_Y$ is massless, no St\"uckelberg couplings of the type $\int_{\mathbb R^{1,3}} F_Y \wedge c_2^\alpha$ or $\int_{\mathbb R^{1,3}} F_Y \wedge c^2_a$  are possible. 
Therefore $F_Y$ can enter the Green-Schwarz terms only via the axionic vertices.
This leaves us with the following possibilities:
The CS term involving $\iota^*C_4$ gives rise to couplings 
\bea \label{AYY1}
{\rm tr}\left(T_Y^2\right) C_S^\alpha   \int_{\mathbb R^{1,3}} c^0_\alpha  \, F_Y \wedge F_Y.
\eea
 In particular no mixed terms of the form $\int_{\mathbb R^{1,3}} F_Y \wedge F_B$ with $F_B$ some other $U(1)$ field strength are possible because ${\rm Tr}\left(T_Y T_B\right)=0$ for all $T_B \neq T_Y$. This axionic vertex combines with the flux-induced St\"uckelberg mass $\int_{\mathbb R^{1,3}} {\rm tr} F_A  \wedge c_2^\alpha$  of some other $U(1)_A$ into the cubic Green-Schwarz term which couples $F_A -  F_Y^2$.

The second type of couplings arise from the CS coupling involving $\iota^*C_2$ and are of the form 
\bea
&&  \int_{\mathbb R^{1,3}}  c^0_a \, F_Y \wedge F_S     \int_S 2 {\rm tr} (T_Y^2)  f_Y \wedge \iota^*\omega_a + \label{YS}  \\
&&  \int_{\mathbb R^{1,3}}  c^0_a \, F_Y \wedge F_Y      \int_S \Big({\rm tr} (T_Y^3)  f_Y +  {\rm tr} (T_Y^2)   f_S\Big)    \wedge \iota^*\omega_a \label{YY}
\eea   
plus a term involving only $F_S$ and $f_S$.
The terms (\ref{YY}) combine with St\"uckelberg couplings $\int_{\mathbb R^{1,3}} {\rm tr} F_A  \wedge c^2_a$ into another set of GS-terms of the form $F_A -  F_Y^2$. Together with the terms from (\ref{AYY1}) they cancel mixed $U(1)_A -U(1)_Y^2$ anomalies. In fact, the cancellation of these anomalies is guaranteed by cancellation of the mixed $U(1)_A -SU(5)^2$ anomalies in models that satisfy the 7-brane and 5-brane tadpole cancellation conditions.

Of primary interest for us are anomalies of the form  ${\cal A}_{U(1)_Y-U(1)^2}$. To cancel these we need the axionic vertex (\ref{YS}), which together with $\int_{\mathbb R^{1,3}} {\rm tr} F_A  \wedge  c^2_a$ induces a GS terms that couples
\bea
   F_A - F_Y - F_S  \;. \label{mixanom}
\eea
Here $F_A$ can be any linear combination of $U(1)$s which is geometrically massive. 
Conversely, mixed anomalies of the type $U(1)_A -  U(1)_B  - U(1)_Y$ must always be of this form as no other GS terms arise. 

It is simple to check this by showing that all $U(1)_A -  U(1)_B  - U(1)_Y$ anomalies where $U(1)_A$ and $U(1)_B$ do not contain a component of $U(1)_S$ vanish. As a result of (\ref{10chir}) the anomaly receives contributions only from the ${\bf 5}$-curves $S \cap D_i$ and $S \cap D_{i'}$.
The anomaly is therefore proportional to
\be
{\cal A}_{U(1)_Y-U(1)_A-U(1)_B} \simeq  \sum_i \left( Q_A^i Q_B^i   \int_S   f_Y  \wedge \iota^* D_i  + Q_A^{i'} Q_B^{i'}  \int_S   f_Y  \wedge \iota^* D_{i'} \right).
\ee
Here the $Q_A^i$ denote the component of the diagonal $U(1)$ associated to the brane wrapping the divisor $D_i$ inside $U(1)_A$, with $Q_A^{i}=-Q_A^{i'}$ being the analogous quantity for the orientifold image. It follows that
\be
{\cal A}_{U(1)_Y-U(1)_A-U(1)_B} \simeq  \sum_i  Q_A^i Q_B^i \int_S   f_Y  \wedge \iota^* (D_i +D_{i'}) \;.
\ee
But since the hypercharge flux had intersection only with the pullback of odd bulk components this vanishes. This clearly applies also to the case $A=B=S$ and so the only anomalies which are relevant are those which mix the $U(1)_S$ and $U(1)_A$. These can be written as
\be
{\cal A}_{U(1)_Y-U(1)_S-U(1)_A} \simeq  \sum_i  Q_A^i      \int_S   f_Y  \wedge \iota^* (D_i -D_{i'})       
\ee
and are indeed cancelled by (\ref{mixanom}).
The important aspect  of this result is that not all such anomalies are cancelled in the field theory. The geometry does impose constraints on the possible hypercharge flux restriction to imply cancellation of a subset of these anomalies, but not all of them! The remaining ones are cancelled by the orientifold odd GS mechanism. 

This result has crucial implications for F-theory model building following from the discussion in section \ref{sec:hypf}. It implies that {\it if} the $U(1)$s in question are the F-theory uplift of the geometrically massive $U(1)$s in IIB, then the constraints imposed in \cite{Palti:2012dd} that the anomalies of type ${\cal A}_{U(1)_Y-U(1)_A-U(1)_B}$ must be cancelled completely in the field theory spectrum can be relaxed. The mechanism which cancels the anomalies that do not vanish in the field theory spectrum is the F-theory uplift of the orientifold odd GS mechanism of type IIB. As yet this uplift is poorly understood, but we have seen that it has an important role to play in model building.

This role gains further importance upon realising that hypercharge flux can also induce new anomalies of type ${\cal A}_{U(1)_Y^2-U(1)_A}$ in the massless spectrum. It is thus possible to  violate the condition (\ref{YY1a}) for geometrically massive $U(1)$s so that (\ref{dp}) can be relaxed. Indeed the only anomalies induced by hypercharge flux and involving geometrically massive $U(1)$s  that can be shown to generically vanish are of type ${\cal A}_{U(1)_Y^2-U(1)_S}$.  This follows from $Q_S^{i}=Q_S^{i'}$, which leads to a coupling $\int_S f_Y  \wedge \iota^* (D_i +D_{i'}) \equiv 0$. Irrespective of this, in the case of geometrically \emph{massless} $U(1)$s (\ref{YY1a}) clearly continues to hold since the GS term cancelling such non-universality in the anomalies (\ref{YY}) vanishes. 

To conclude this section we turn to a related phenomenon also pointed out in \cite{Palti:2012dd},  where it was shown that similar constraints arise in the presence of $U(1)$ fluxes other than hypercharge charge as long as these do not induce a St\"uckelberg mass term. In this case the relevant anomalies are pure GUT ones of $SU(5)^2-U(1)$ type. We now show that a similar resolution for these anomalies also applies. Consider first turning on flux $f_i$ along a $U(1)_i$ which is the diagonal $U(1)$ of a brane wrapping the divisor $D_i \neq S$. Then the chiral spectrum of $\f$ states is
\bea
\f_{-1_i}\;&:&\;   -  \int_{D_i} f_i \wedge i^* S  =  - \half \int_{D_i} f_i \wedge i^*\left( S - {S'} \right) \;, \nn \\
\f_{+1_i}\;&:&\;   - \int_{D_i} f_i \wedge i^* {S'}  = \half \int_{D_i} f_i \wedge i^*\left( S - {S'} \right) \;, \label{oddu1flux}
\eea
where  we have used the fact that by assumption the flux couples only to the odd components in order to induce no St\"uckelberg mass term. This spectrum has no net contribution to the anomaly $SU(5)^2-U(1)_i$ because the two types of $\f$s contribute equally but have opposite charge under $U(1)_i$. This is consistent with the fact that there can be no Green-Schwarz contribution to this anomaly from such a flux since the flux would have to appear in the same trace as over the external $SU(5)$ generators; this, however, is not possible since the latter arise from different branes. The flux does appear to induce a pure non-Abelian anomaly though, which must be cancelled once D7-tadpole cancellation (\ref{D7tad}) is imposed. To see this we intersect the D7-tadpole with $f_i \wedge \left(S-{S'}\right)$ and use $f_i \wedge S \wedge {D}_{i'} = -f_i \wedge {S'} \wedge D_i$ to obtain
\be
5 f_i \wedge S \wedge S  + f_i \wedge \left(S - {S'}\right)\wedge \sum_j  D_j = 0 \;.
\ee
The second term is nothing but the chirality induced by such a flux as in (\ref{oddu1flux}), and as discussed around (\ref{YSS0})  in absence of exotic ${\bf 15}$-curves the first term vanishes. We therefore find that depending on the self-intersection of $S$ either no chirality can be induced or if it does there are additional states that cancel the anomaly. 

What remains is to consider flux along $U(1)_S$. Unlike the other fluxes this can induce a mixed anomaly of type $SU(5)^2-U(1)_A$, but  the trace structure does allow for such anomalies to be cancelled by the Green-Schwarz mechanism. It is straightforward to check this cancellation, see \cite{Plauschinn:2008yd} for a general calculation.

\subsection{A family of  $SU(5) \times U(1)$ models}
\label{sec_311}	

In this section we present a  model prototypical of the general constructions we have been studying. Our brane setup gives a direct realisation of matter curves with orientifold odd components and so supports net hypercharge flux restriction to the matter curves. We consider the brane configuration
\bea \label{setup1}
5 (S + S'), \qquad D_1 + D_{1'}, \qquad D_2 + D_{2'}.
\eea
Tadpole cancellation (\ref{D7tad}) enforces that
\bea 
D_1    &=&  4 O7 - a S - (5-a)  S' - b D_2 - (1-b)  D_{2'}, \nonumber  \\
D_{1'} & =& 4 O7 - (5-a) S - a  S' - (1-b) D_2 - b  D_{2'}. \label{hom-rel1}
\eea
This is the most general solution if we assume that $D_1$ has no contribution from negative classes other than $S-S'$ and $D_2- D_{2'}$. Such extra contributions would lead to a geometric mass matrix $M_I^a$ (\ref{Massmatrix}) of full rank and thus allow for no geometrically massless $U(1)$.
 If there are  no further homological relations between the divisors,
the only geometrically massless combination of diagonal $U(1)$s is given by
\bea
U(1)_X = - \frac{1}{2} \Big( (5-2a) \, U(1)_S - 5 U(1)_1 + (5-10b) U(1)_2    \Big),
\eea
where we have chosen a convenient normalization.
In principle, every choice of $a$ and $b$ gives rise to a different class of $SU(5) \times U(1)$ models provided one can find suitable \emph{holomorphic} divisors $S$ and $D_i$ satisfying (\ref{hom-rel1}) on a given Calabi-Yau three-fold $X$ and orientifold projection.
As an example, for $a = 2, b= 2$  the massless Abelian gauge group is the combination
\bea
U(1)_X =  - \frac{1}{2} \Big(U(1)_S - 5 U(1)_1 - 15 U(1)_2 \Big). \label{u1x}
\eea
The described setup gives rise to the spectrum summarized in table \ref{tab:spec1}.

It is not hard to construct explicit realizations of this brane setup for a concrete compactification Calabi-Yau $X$ and orientifold involution, e.g.\ along the lines of \cite{Blumenhagen:2008zz},
but we do not present such a construction here as it would not add much new to our general understanding. 

\begin{table}
\centering
\begin{tabular}{|l|c|c|c || c|  c | c | c |   }
\hline
\makebox[8mm]{Curve} & Locus & $q_X$ & \parbox[c][10mm][c]{16mm}{\center  \vspace*{-4mm}State\\$a=b=2$} &   \makebox[8mm]{Curve} & Locus & $q_X$  &\parbox[c][10mm][c]{16mm}{\center  \vspace*{-4mm}State $a=b=2$ } \\
\hline
$C_{\bf 10}$ & $S \cap O7$ & $ 2a-5 $ & ${\bf 10}_{-1}$ &$ C_{\bf 1^{(1)}}$ & $ D_1 \cap D_{1'}$  & $ 5 \,$ & ${\bf 1}_{\pm 5}$\\
\hline
$C_{\bf 5^{(1)}}$ & $ S \cap D_1$      &  $ a - 5$    &   $  {\bf 5}_{-3}$  & $ C_{\bf 1^{(2)}}$ & $ D_2 \cap D_{2'}$  & $ 5-10b \,$ &  ${\bf 1}_{\pm 15}$ \\ 
\hline
$C_{\bf 5^{(2)}}$ & $ S \cap D_{1'}$   & $ a$  &  $  {\bf 5}_{+2}$ & $C_{\bf 1^{(3)}}$  & $ D_1 \cap D_2$      &  $5-5b \,  $      & $ {\bf 1}_{\pm 5}$\\
\hline
$C_{\bf 5^{(3)}}$ & $ S \cap D_2$      &  $ a - 5b$  &  $ {\bf 5}_{-8}$ & $  C_{\bf 1^{(4)}}$ & $ D_1 \cap D_{2'}$ & $ 5b\, $ &  ${\bf 1}_{\pm 10}$ \\
\hline
$C_{\bf 5^{(4)}}$ & $ S \cap  D_{2'} $  &   $ a+5b - 5$   &   $ {\bf 5}_{7}$   &&&& \\
\hline
\end{tabular}
\caption{Spectrum of $SU(5) \times U(1)$ model with 7-branes.}
\label{tab:spec1}
\end{table}

As pointed out already, the ${\bf 5}$-matter curves have orientifold even and odd components and thus hypercharge flux can consistently restrict to the matter curves as required to realize doublet-triplet splitting. 
As a cross-check, the mixed $(U(1)_X)^2 - U(1)_Y$  and  $U(1)_X - (U(1)_Y)^2$ anomalies vanish. Since $\int_{C_{\bf 10}} f_Y =0$ because of (\ref{orthY}), the anomalies are proportional to $\sum_a q^2_X({\bf 5}^{(a)}) \int_{C_{\bf 5^{(a)}}}   f_Y$ and to $\sum_a q_X({\bf 5}^{(a)}) \int_{C_{\bf 5^{(a)}}}   f_Y$, respectively. Both sums vanish because as a consequence of (\ref{hom-rel1}) and (\ref{orthY}) the integral over the matter curves are related as
\bea
&& \int_{C_{\bf 5^{(1)}}}   f_Y =  - \frac{2b-1}{2} \int_S f_Y \wedge \iota^*(D_2 - D_{2'}) = - \int_{C_{\bf 5^{(2)}}}   f_Y, \\
&& \int_{C_{\bf 5^{(3)}}} f_Y =   \frac{1}{2} \int_S f_Y \wedge \iota^*(D_2 - D_{2'}) =  - \int_{C_{\bf 5^{(4)}}} f_Y. \label{hyperresmod1}
\eea
Note that we have used $\int_{S} f_Y \wedge  \iota^* S = \int_{S} f_Y  \wedge \iota^*S' = 0$, which follows from imposing absence of exotic ${\bf 15}$-matter curves (see the discussion around (\ref{YSS0})).

The family of brane setups (\ref{setup1}) is the Type IIB version of a number of $SU(5) \times U(1)$ models that have featured prominently in the recent F-theory literature.
The special choice $a=b=2$ corresponds to the charge assignments of the model presented in \cite{Braun:2013yti}. The appearance of orientifold odd matter curve components in Type IIB makes us confident that also in F-theory this class of models allows for a non-trivial restriction of hypercharge flux to the ${\bf 5}$-curves, even though this remains yet to be shown. 
If one replaces the 7-brane pair along $D_2 + D_{2'}$ by a single Whitney brane $W$, the spectrum reproduces what is called in the F-theory literature the $SU(5) \times U(1)_{PQ}$ model with  three ${\bf 5}$-curves in the version  with only a  single ${\bf 10}$-curve. This model was introduced in \cite{Marsano:2009wr} in a local split spectral cover and realized in a Calabi-Yau four-fold in \cite{Mayrhofer:2012zy}. 
Note that after replacing $D_2 + D_{2'}$ by $W$ the orientifold odd components of the ${\bf 5}$-matter curves vanish, in agreement with the conclusions of \cite{Palti:2012dd} that no consistent hypercharge restrictions to the matter curves are possible in the $SU(5) \times U(1)_{PQ}$ with a single ${\bf 10}$-curve.
If the brane pair $D_2 + D_{2'}$ is removed altogether, we arrive at the $SU(5) \times U(1)_X$ model with a single ${\bf 10}$-curve and two ${\bf 5}$-curves studied in \cite{Marsano:2009gv}  and  \cite{Krause:2011xj,Grimm:2011fx,Krause:2012yh}. Similarly, one can add more brane-image brane pairs to arrive at even more ${\bf 5}$-curves.

\subsubsection*{A $U(1)_{PQ}$ model without exotics}

It is important to point out that although (\ref{dp}) holds for the case of $U(1)_X$ and also $U(1)_S$, matching the general discussion in section \ref{sec:iibhypano}, it does not hold for all the $U(1)$s. For example in the cases of $U(1)_1$ and $U(1)_2$ it manifestly fails to hold since the states on $C_{\bf 5^{(1)}}$  and $C_{\bf 5^{(2)}}$ have opposite flux restriction but also opposite charges. The resulting non-universality in the anomalies is canceled by the contribution coming from (\ref{YY}). This opens up a way to bypass a problem highlighted in \cite{Dudas:2010zb} which is that the constraints (\ref{dp}) imply that the presence of a $U(1)_{PQ}$ symmetry, when combined with hypercharge flux doublet-triplet splitting, leads to exotics in the massless spectrum. Let us present a toy example of how this result is evaded. Consider modifying the solution (\ref{hom-rel1}) by allowing for another arbitrary odd homology class $\chi$
\bea 
D_1    &=&  4 O7 - a S - (5-a)  S' - b D_2 - (1-b)  D_{2'} + \chi, \nonumber  \\
D_{1'} & =& 4 O7 - (5-a) S - a  S' - (1-b) D_2 - b  D_{2'} - \chi. \label{hom-rel1chi}
\eea
This leaves the D7-tadpole invariant. The first implication of this is that all the $U(1)$s are now geometrically massive. The fact that all the hypercharge flux restrictions in (\ref{hyperresmod1}) are proportional can be understood from the requirement that the hypercharge induce no anomalies in the massless spectrum with respect to the massless $U(1)$ in the model. Having made all the $U(1)$s massive we expect more freedom since anomalies in the massless spectrum can be canceled by (\ref{YY}). Indeed if we choose 
\be
\int_S f_Y \wedge \iota^*(D_2 - D_{2'}) = 0 \;,
\ee
we now have
\bea
&& \int_{C_{\bf 5^{(1)}}}   f_Y =  \int_S f_Y \wedge \iota^*\chi = - \int_{C_{\bf 5^{(2)}}} f_Y, \\
&& \int_{C_{\bf 5^{(3)}}} f_Y   =  \int_{C_{\bf 5^{(4)}}} f_Y = 0. \label{hyperresmod1}
\eea
Therefore we can induce doublet-triplet splitting on $C_{\bf 5^{(1)}}$ and $C_{\bf 5^{(2)}}$ with no other non-GUT exotics present. The matter curves $C_{\bf 5^{(1)}}$ and $C_{\bf 5^{(2)}}$ are not vector-like with respect to (massive) $U(1)$ combinations, for example (\ref{u1x}). Therefore we have an effective global $U(1)_{PQ}$ symmetry, which can only be broken non-perturbatively, and doublet-triplet splitting with no implied exotics.

\section{Hypercharge flux and Gauge Coupling Unification}
\label{sec:hypgauge}

In this section we study some effects of hypercharge flux on gauge coupling unification. We will work primarily in the IIB framework, much in the spirit of \cite{Blumenhagen:2008aw}, but taking into account the new hypercharge effects discussed in the previous section.  The terms of interest for us are the following couplings  appearing in the CS action (\ref{SCS1}), 
\be
S_{CS} \supset \frac{1}{2} \mu_7  \int_{D7} i^*C_4 \wedge {\rm tr }\left[{F} \wedge {F}\right] + \frac{1}{6} \mu_7  \int_{D7} i^*C_2 \wedge {\rm tr }\left[ {F} \wedge {F} \wedge {F}\right] + \frac{1}{24} \mu_7  \int_{D7}C_0  \;{\rm tr }\left[ {F} \wedge {F} \wedge {F} \wedge {F}\right] \;. \label{csf4f3}
\ee
Note that the first term also has a contribution from the NS B-field, which we drop for ease of notation for now. It can be easily reinstated as it appears always in the combination $C_2 - C_0 B$.  The above CS terms include the contributions
\be
S_{CS} \supset \frac12 \int_{\mathbb R^{1,3}} c_0^a \;{\rm tr }\left[ F \wedge F \int_{S} i^*\omega_a \wedge f\right] + \frac14 \int_{\mathbb R^{1,3}} C_0 \;{\rm tr }\left[ F \wedge F \int_S f \wedge f\right] \;. \label{d7cscorr}
\ee
These contributions can be used to calculate the {tree-level} effects of flux on the gauge couplings in Type IIB \cite{Blumenhagen:2008aw} since by four-dimensional supersymmetry they appear together in the gauge kinetic function. We consider turning on internal flux along the generators
\be
f = 
f_S \, \mathrm{diag}\left(1,1,1,1,1\right) + \frac15 f_Y \, \mathrm{diag}\left(-2,-2,-2,3,3\right)\;. \label{normflux}
\ee
Note that our normalisation of the internal flux differs by a factor of $\frac{1}{6}$ from that of the external field strength in (\ref{norm4D}).
It is convenient to define the quantity
\be
\tilde{f}_S \equiv f_S - \frac25 f_Y \;. \label{tf0}
\ee
We decompose the external field-strengths in terms of the commutants of $U(1)_Y$ within $SU(5)$ as 
\be
F = F_{SU(3)}^a T^{SU(3)}_a + F_{SU(2)}^i T^{SU(2)}_i + F_Y T_Y + F_S T_S^0 \;.
\ee
In this notation we find for (\ref{d7cscorr}) the expression
\bea
\frac14 \left( {\rm tr}\left(F^2_{SU(3)}\right) N  
+ {\rm tr}\left(F^2_{SU(2)}\right) \left[N + M\right] 
+ \frac56 F_Y^2 \left[N + \frac35M\right] 
+ 5 F_S^2 \left[N + \frac25 M\right] 
+ F_S F_Y M \right) \; \label{corrfull}
\eea
with
\bea
N =  \int_S \tilde{f}_S \wedge \left( C_0\tilde{f}_S + 2 i^*C_2 \right),  \qquad  M = \int_S f_Y \wedge \left(2 C_0\tilde{f}_S + C_0f_Y + 2i^* C_2 \right).
\eea

There is a lot of interesting physics in the expression (\ref{corrfull}), which we outline below.
The internal flux induces kinetic mixing between $F_S$ and $F_Y$. This can be undone by an appropriate shift of $F_Y \rightarrow  F_Y + \alpha F_S$. This shift retains the appropriate normalisation for the hypercharge, which is important for gauge coupling unification. Note that this is the appropriate shift, rather than a general rotation mixing the hypercharge and the diagonal $U(1)_S$, because it leaves the mass term for $U(1)_S$ invariant. It also has the interesting effect of charging the SM fields under the diagonal $U(1)$ by some fractional charge, in a non-GUT universal way. Since the $U(1)$ in this case is very massive the phenomenological implications of this are not drastic, though it would be interesting to study this in more detail. Note that the kinetic mixing is present already from just the second term of (\ref{d7cscorr}) which means it can be induced even if the hypercharge does not couple to $i^* C_2$. In fact in this case its magnitude is fixed by the requirement of the absence of bulk exotics, at least if $S$ is taken to be a del Pezzo surface, \cite{Beasley:2008kw,Donagi:2008kj}
\be
\int_S f_Y \wedge f_Y = -2 \;. \label{fynoex}
\ee
This implies significant mixing. It is interesting to note that in the case where the diagonal $U(1)$ is not anomalous, but still massive, its mass can be quite low for large values of the volume, such as in the LARGE volume scenario \cite{Goodsell:2009xc} (though this is at odds with gauge coupling unification at the usual GUT scale). 

The aspect of (\ref{corrfull}) that we are most interested in at this point is its effect on gauge coupling unification. We see that the tree-level expression for the non-GUT universal contribution to the real part of the gauge kinetic functions is 
\be
\delta{{\rm Re}f_i} =  \frac12 \delta_i \int_S f_Y \wedge \left[C_0 \left(2 \tilde{f}_S +  f_Y \right) + 2i^* C_2 \right] \;, \label{ganonuni}
\ee
with $\delta_{SU(3)}=0$, $\delta_{SU(2)}=1$, $\delta_{U(1)_Y}=\frac35$.\footnote{Note that the $\delta_i$'s correspond to the contribution to the beta functions from a Higgs-type doublet $(1,{\bf 2})_{1/2}$ (or equivalently its $SU(5)$ triplet partner). This matches the results of \cite{Conlon:2009kt} where in the context of D3-branes at singularities the (N=1) twisted modes coupled in the gauge kinetic function proportional to the $\beta$-function contribution of the associated open string subsector. Indeed the modes $C_2$ here are precisely the large volume versions of these twisted modes and the corresponding open string sector is the matter on the $\f$-matter curves induced by the hypercharge flux. Note also that since the hypercharge flux can couple to them, this presents a slight modification of the picture presented in \cite{Conlon:2009qa} in the IIB case where the hypercharge was taken to be trivial, however the running to the scale of the associated 'tadpole' remains. In the F-theory picture the hypercharge truly is globally trivial and the conclusions of \cite{Conlon:2009qa} hold more precisely.}

 If we set $\tilde{f}_S=i^* C_2=0$ we still find a tree-level splitting of the gauge couplings purely due to hypercharge and the constraint (\ref{fynoex}). This was first pointed out in \cite{Blumenhagen:2008aw}. 
The splitting can be estimated to be of order 4-5\%: At the GUT scale the inverse gauge couplings take the value $\frac{4\pi}{g^2}\sim24$ and this is the imaginary part of the gauge kinetic function, the real part being $C_4$, and therefore the splitting is relative to the $C_4$ coefficient in (\ref{csf4f3}). Furthermore we have $\delta_{SU{3}}-\delta_{SU{2}} = 1$ and expect the dilaton superpartner to $C_0$ to have an order one vev in F-theory. Together with the factor of $2$ from (\ref{fynoex}) the gauge coupling splitting thus goes like $24 \pm 1$. Such a splitting is uncomfortably large when it comes to gauge coupling unification which in the MSSM is of order 3\% at two loops. It is important to emphasise that this is really an estimate based exclusively on the tree-level contribution to the gauge kinetic function and a naive extrapolation to F-theory of this IIB result. In a full account this splitting of the gauge coupling receives further corrections. One source of these are the usual field theory corrections due to threshold effects from massive Landau-levels, which are expected to be small at least if the massive states are close to the GUT scale. Another source of corrections arises because (\ref{csf4f3}) is not the full expression for the $F^4$ terms that can arise in IIB and F-theory. This tree-level piece is corrected at 1-loop giving a schematic term $\mathrm{ln} \tau F^4$, and by $D(-1)$ instantons leading to terms of type $e^{i\tau} F^4$ (see \cite{Billo':2010bd} for recent advances on calculating these). Here $\tau = C_0 + i s$ represents the axio-dilaton superfield. These latter corrections are subdominant to (\ref{ganonuni}) in the weak-coupling limit $s \rightarrow \infty$. However at strong coupling they can compete and alter the splitting. In order to be able to estimate this a much better understanding of the origin of the $F^4$ terms in F-theory is needed. Some work along these lines has utilised Heterotic/F-theory duality \cite{Donagi:2008kj} (see also \cite{Dolan:2011aq}) and M-theory warping \cite{Grimm:2012rg}, though a detailed quantitative analysis of these corrections for realistic F-theory models is still missing. On general grounds however it is reasonable to require that the IIB tree-level correction which we study here should not give too large a splitting of the gauge coupling. If it does then it is possible that in fully-fledged realistic F-theory models all the other effects and corrections conspire to cancel this large splitting, though in the absence of an explicit computation showing this one would prefer to not rely on such a possibility. Fortunately, already in IIB it is clear that the tree-level splitting (\ref{ganonuni}) has additional structure further to the simplest case just discussed which can change the conclusions regarding gauge coupling splitting.

The first modification we would like to study is to consider $\tilde{f}_S\neq0$ but still $i^* C_2=0$. Already in \cite{Blumenhagen:2008aw} it was suggested that if we turn on  {\it globally trivial}	$\tilde{f}_S$ it can be arranged for $\int_S f_Y \wedge \tilde{f}_S = 1$, which would the nullify the splitting (\ref{ganonuni}). The requirement of global triviality was presumably imposed because it was thought that no globally non-trivial elements can have a non-vanishing intersection with the hypercharge flux. The first new observation we can make is that using our current results we see that this is actually not a necessary requirement because $\tilde{f}_S$ can have components in the image of the pullback of the $H^{1,1}_-(X)$.

There is another interesting observation we can make regarding $\tilde{f}_S$ which relates to its use rather than that of $f_S$ in (\ref{corrfull}). Recall the two are  related by a shift in the hypercharge flux (\ref{tf0}). The reason was given in \cite{Beasley:2008kw,Donagi:2008kj} and relates to the quantisation of the flux. The important point is that for all states in the theory sensitive to some combination of flux, the flux appearing in the Dirac equation of the state should be integer quantised. Let us consider the states in the theory and the flux combinations which they feel. We turn on the flux $\frac15 f_Y$ along the hypercharge generator normalised as in (\ref{normflux}), $f_S$ along the diagonal $U(1)$, and $f_i$ along the $U(1)$ branes intersecting $S$ along the {\bf 5}-matter curves. Then the matter states couple to the gauge bundles whose first Chern classes are given by the following combination of fluxes,
\bea
\left({\bf 3},{\bf 2}\right)_{-5/6}\;&:&\; -f_Y \;, \nn \\
\left({\bf 3},{\bf 2}\right)_{1/6}\;&:&\; \frac15 f_Y + 2f_S \;, \;\;
\left({\bf \bar{3}},1\right)_{-2/3}\;:\; -\frac45 f_Y + 2f_S \;, \;\;
\left(1,1\right)_{1}\;:\; \frac65 f_Y + 2f_S \;, \nn \\
\left({\bf 3},1\right)_{-1/3}\;&:&\; -\frac25 f_Y + f_S - f_i \;, \;\;
\left(1,{\bf 2}\right)_{1/2}\;:\; \frac35 f_Y + f_S - f_i \;.
\eea
These bundles are required to be integer quantised\footnote{We ignore here the possible overall shift due to the Freed-Witten anomaly.} for all the states present in the theory.
A natural solution that guarantees this is to write $f_S=\tilde{f}_S + \frac25 f_Y$, where now $f_Y$, $\tilde{f}_S$, and $f_i$ are integer quantised. This is the reason why $\tilde{f}_S$ is the natural integer quantised object to work with. Let us denote this twisting of the diagonal $U(1)$ flux as twisting of type 1. 

Suppose now that the hypercharge bundle $L_Y$ restricts trivially to the $\te$-curve. This means that the Poincar\'e dual of $f_Y \in H^2(S)$ does not have any intersection with the $\te$-matter curve. This happens, for example, whenever the hypercharge flux is purely orientifold odd because then it vanishes pointwise over the $\te$-matter curve, since the latter lies on top of the orientifold.
In this situation we can also take a solution $f_i=\tilde{f}_i-\frac25 f_Y$ where now $\tilde{f}_i$, $f_S$ and $f_Y$ are integer quantised. We denote this as twisting of type 2. Note that no problems occur for the GUT singlets because these arise at intersections away from the GUT brane and therefore again the hypercharge flux vanishes geometrically when restricted to such loci.
In this case the splitting of the gauge kinetic function is most conveniently written (still for vanishing $\iota^* C_2$) as
\be
\delta{\rm Re}f_i =  \frac{1}{10} \delta_i C_0 \int_S f_Y \wedge f_Y  +  \delta_i C_0 \int_S f_Y \wedge f_S      \; \label{split2}.
\ee
If in addition we arrange for $\int_S f_Y \wedge f_S =0$, e.g.\ by ensuring that $f_S$ is in the pullback of $H^{1,1}_+(X)$, the only contribution is from the first term.
But this is a factor of 5 smaller than the case with $\tilde{f}_S=0$ and gives an estimate of only a 1\% correction, which is well within the MSSM 2-loop result.

In fact  in F-theory compactifications  the twisting of type 2 is easily implemented to the extent that it amounts to considering a suitably quantised linear combination of $G_4$ fluxes associated with hypercharge and with the massless $U(1)$s of the model, see (\ref{sec_hyperG4}) for details. 
For example consider the set of models corresponding to the brane setup of section \ref{sec_311}. It is simple to check that a single twisting of the $U(1)$ flux makes the hypercharge flux for all the matter curves integer. This is guaranteed by the fact that the charges of the matter curves under the $U(1)$ differ by multiples of 5.

So far we have discussed the gauge coupling splitting in the cases where $\int_S f_Y \wedge i^* C_2  = 0$. However, if $f_Y$ has net restriction to the matter curves the situation is more complicated because then necessarily $\int_S f_Y \wedge i^* C_2  \neq 0$. In such a case gauge coupling unification depends on the vacuum configuration of the moduli. Let us recall some of the structure of the orientifold odd moduli sectors, following the discussion in \cite{Grimm:2011dj}. 
In terms of the cohomology decomposition on $X$ introduced before (\ref{expansion1})  the K\"ahler form $J$ of $X$ and the NS-NS form have an expansion 
\bea
J &=& v^{\alpha} \omega_{\alpha}, \quad\quad  C_2 = c^{a} \omega_a \;, \\
B_2 &\equiv&   B_+ + B_- = b^{\alpha} \omega_{\alpha} + b^{a} \omega_a \;.
\eea
Here  $b^{\alpha}$  can only take the discrete values $0$ or $ \frac{1}{2}$ consistent with the orientifold action. The appropriate chiral fields for these compactifications are given by \cite{Grimm:2004uq,Jockers:2004yj}
\bea
G^a &=& c_0^{a} - \tau b^a,  \label{GaTalpha} \\
T_{\alpha} &=& \frac{1}{2}\kappa_{\alpha \beta \gamma} v^\beta v^\gamma+ i \left(c^0_{\alpha} - \kappa_{\alpha bc} c_0^b {b}^c \right) + \frac{i}{2} \tau \kappa_{\alpha b c}  b^b   \, b^c \;, \nonumber
\eea
where we have defined the intersection numbers
\bea
\kappa_{\alpha \beta    \gamma} =  \int _{X}  w_{\alpha}    \wedge     w_{\beta} \wedge w_{\gamma}, \qquad 
\kappa_{\alpha b    c} &=& \int _{X}     w_{\alpha}    \wedge  w_{b} \wedge w_{c} \;.
\eea
The relevant moduli for the splitting (\ref{ganonuni}) are the $b^a$. In order to work with the four-dimensional superfields it is convenient to define the quantities
\be
p_{YY} \equiv \int_S f_Y \wedge f_Y \;,\;\; p_{YS} \equiv \int_S f_Y \wedge \tilde{f}_S \;,\;\; p_{Ya} \equiv \int_S f_Y \wedge i^*\omega_a \;.
\ee
 We can then write the splitting of  ${\rm Im}f_i = \frac{4 \pi}{g_i^2}$ as
\be
\delta {\rm Im}f_i = \frac12 s \delta_i \left(2 p_{YS} + p_{YY} - 2 b^a p_{Ya}\right) \;. \label{ggsmod}
\ee
In order to determine the vacuum expectation values of the $b^a$ we should consider the two sources of potentials for them. The first is the D-term contribution associated to the diagonal $U(1)$ of a brane wrapping the divisor $D_I$ carrying flux along the $U(1)$ of $f_I=f^a_I \omega_a+f^{\alpha}_I \omega_{\alpha}$, which is given by \cite{Grimm:2011dj}
\bea
  D_I = \frac{\ell_s^2}{4  \pi    \mathcal{V}} v^\alpha \big   (\kappa_{\alpha bc} (b^b - f^b_I) C^c_I -\kappa_{\alpha \beta \gamma} f^\beta_I C^\gamma_I\big) \ \nn 
\eea
with ${\cal V}$ the Calabi-Yau volume. Note that we have not displayed any charged matter fields which have to be added to this supergravity contribution. Also the even fluxes here are the appropriate combination of flux and $b^{\alpha}$ which are integer quantised. The other potential contribution for the $b^a$ comes from fluxed instantons in the superpotential as studied in \cite{Grimm:2011dj}. We do not go into the details here and refer to \cite{Grimm:2011dj} for the appropriate expressions. The determination of the vev of the $b^a$ is clearly a model dependent question, and it is interesting to see this explicit connection between gauge coupling unification and moduli stabilisation. It would be illuminating to study explicit models where all the fluxes and intersection numbers are specified and to determine the effect on gauge coupling unification.

\section{Hypercharge $G_4$-flux in F-theory}
\label{sec_hyperG4}

A lot of progress has been made recently in the explicit description of $G_4$ gauge fluxes that do not break the $SU(5)$ symmetry \cite{Braun:2011zm,Marsano:2011hv,Krause:2011xj,Grimm:2011fx}, 
but the analysis of hypercharge flux in the local model building literature has been mostly in the language of two-form flux $f_Y$ along the $SU(5)$ divisor as inspired by the Type IIB picture (see section \ref{sec:hypf}). 
In this section we give a definition of hypercharge flux directly in terms of $G_4$-flux defined by a class in $H^{2,2}(\hat Y_4)$ of the fully resolved Calabi-Yau four-fold $\hat Y_4$. This includes the construction of hypercharge flux in terms of the four-form classes dual to the matter surfaces of the resolved Calabi-Yau $\hat Y_4$. We will also briefly discuss  the twisting pertinent to the quantization of hypercharge flux.

The four-dimensional gauge potential $A_Y$ associated with hypercharge arises via F/M-theory duality by expanding the M-theory three-form $C_3$ in terms of $\tw_Y \in H^{1,1}(\hat Y_4)$ of the resolved Calabi-Yau four-fold $\hat Y_4$,
\bea \label{defwy}
C_3 = \frac{1}{6} A_Y \wedge \tw_Y+ \ldots, \qquad \quad \tw_Y = \sum_{i=1}^4 l_i E_i \qquad l_i=(-2,-4,-6,-3).
\eea
Here $E_i \in  H^{1,1}(\hat Y_4)$ denote the two-forms dual to the exceptional divisors $e_i$, $i=1, \ldots, 4$ introduced in the process of resolving the $SU(5)$ singularity in the fiber over the GUT brane.  Their  intersection numbers
\bea
\int_{\hat Y_4} E_i \wedge E_j \wedge \pi^*D_a \wedge \pi^* D_b = C_{ij}\int_B S \wedge D_a \wedge D_b \qquad \forall \, D_a \in H^{1,1}(B)
\eea
involve the Cartan matrix $C_{ij}$ of $SU(5)$, with the convention that $C_{ii}=-2$. The factor of $\frac{1}{6}$ is chosen such that the $U(1)_Y$ charges comply with the conventions in (\ref{norm4D})
Each divisor $e_i$ is $\mathbb P^1$-fibration over $S$, and we denote the fiber by $\mathbb P^1_i$.
The above definition of $\tw_Y$ ensures that 
\bea \label{wYe3}
\int_{\mathbb P^1_i} \tw_Y = 5 \, \delta_{i\, 3}.
\eea
Note that this description is of course very  similar to the analysis in \cite{Donagi:2008kj} of hypercharge $G_4$-flux in the language of a local ALE-fibration over the GUT brane $S$.

The internal hypercharge flux is described by an element $G_4 \in H^{2,2}(\hat Y_4)$ which breaks $SU(5) \rightarrow SU(3) \times SU(2) \times U(1)_Y$. In order to correspond to a gauge flux it is subject to the usual transversality constraint
\bea \label{transverse}
\int_{\hat Y_4} G_4 \wedge \pi^*D_a \wedge \pi^* D_b = 0  = \int_{\hat Y_4} G_4 \wedge Z \wedge \pi^* D_a.
\eea
To work out the requirements for $G_4$ to break the $SU(5)$ we 
consider the pullback of $G_4$ to the divisor $e_i$, take its dual two-cycle and use the projection map to push it forward to the base. Because of~\eqref{transverse} the 2-cycle remains a two-cycle under the projection.
The final two-form is defined as the dual to this projected curve on the base of $e_i$, which is just the GUT four-cycle $S$ on $B$.
Sloppily we refer to this operation as integrating $G_4$ over $\mathbb P^1_i$ such as to produce an element in $H^{1,1}(S)$.
With this understanding 
 the group theoretic condition for hypercharge breaking thus becomes, in analogy with (\ref{wYe3}),
\bea \label{hyperG1}
\int_{\mathbb P^1_i} G_4 = 0 \qquad i = 1,2,4, \qquad \quad \int_{\mathbb P^1_3} G_4 = f_Y, \quad f_Y \in H^{1,1}(S).
\eea
The two-form $f_Y \in H^{1,1}(S)$ is what is usually called hypercharge flux in Type IIB inspired  7-brane language. Note that the factor of $5$ from (\ref{wYe3}) has been absorbed in $f_Y$.

We next turn to the condition for absence of St\"uckelberg masses. From the gauged supergravity analysis \cite{Donagi:2008kj,Grimm:2010ks,Grimm:2011tb} the masslessness constraint is that
\bea \label{hyperG2}
\int_{\hat Y_4} G_4 \wedge \tw_Y \wedge \pi^*D_a = 0 \qquad \forall \, D_a  \in H^{1,1}(B).
\eea
The wedge product with $\tw_Y$ can be worked out with the help of  (\ref{wYe3}) to yield
\bea
\int_S f_Y \wedge \iota^*D_a = 0 \qquad  \forall \,  D_a   \in H^{1,1}(B),
\eea
which is precisely the constraint that $\iota_! f_Y =0$  \cite{Donagi:2008kj}. Note that this constraint holds on the base $B$ of the elliptic fibration.

An explicit construction of hypercharge $G_4$ is possible as follows:
First one can simply consider the 4-cycle ${\cal C}_{A i}$ defined by fibering any of the $\mathbb P^1_i$ over an arbitrary curve $C_A$ in $S$ on $B$. This is nothing but the restriction of  $e_i$ to $C_A$, ${\cal C}_{A i} = e_i|_{C_A}$. The dual four-form is denoted by $[{\cal C}_{A i}] \in H^{2,2}(\hat Y_4)$. Any linear combination 
\bea
G_4 = \sum_{A,i} l_i [{\cal C}_{A i}], \qquad l_i=(-2,-4,-6,-3)
\eea
automatically satisfies (\ref {transverse})  and thus defines a hypercharge flux in the above sense if 
\bea
 \sum_A \int_S   [C_A] \wedge \iota^*D_a = 0 \qquad \forall \, D_a \in H^{1,1}(B)
\eea
holds.
Note in particular that $ f_Y = \sum_A  [C_A]$. This description coincides with the form of hypercharge fluxes given in the local ALE-analysis of \cite{Donagi:2008kj}.

In addition we now  construct a hypercharge flux in terms of the four-forms dual to the matter surfaces which appear in the process of the $SU(5)$ resolution  \cite{Esole:2011sm,Marsano:2011hv,Krause:2011xj,Grimm:2011fx}.
Let us denote by $C_a$ the curve on $S$ on which matter states in the representation $R_a$ of $SU(5)$ are localised. 
In the resolved fiber over $C_a$, one or more of the $\mathbb P^1_i$ split. 
The resolved fiber over a generic point on $C_a$ degenerates into a tree of ${\mathbb P}^1_I$ (with certain multiplicities) labelled by an index $I$. 
Let us denote the 4-cycle given by fibering $\mathbb P^1_I$  over $C_a$ by  $Z_{aI}$ with dual four-form class  $[Z_{aI}] \in H^{2,2}(\hat Y_4)$.
To the extent that not all of these $Z_{aI}$ are of the form $e_i|_{C_a}$ the use of such 4-cycles for $G_4$-flux goes beyond the above approach.
To each of these $Z_{aI}$ one can associate a vector $v_{aI}[i]$ defined such that
\bea
\int_{\mathbb P^1_i}  [Z_{aI}] =  v_{aI}[i]  [C_a]  \in H^{1,1}(S), \qquad i=1,2,3,4,
\eea
where $[C_a]$ is the two-form dual to the curve $C_a$.
For completeness let us recall how to construct out of these the matter surfaces: To each representation $R_a$ we consider the associated collection of weight vectors $\beta^{(n)}$, $n=1, \ldots, {\rm dim}(R_a)$. Each $\beta^{(n)}$ is a 4-vector and can be written as a linear combination of $ \beta^{(n)} = \sum_I \alpha^{(n) I} v_{aI}[i]$.
This defines a collection of surfaces ${\cal C}^{(n)}_{a} = \sum \alpha^{(n) I} Z_{aI}$. The states
 correspond to M2-branes wrapping the associated linear combinations of $\mathbb P^1_I$ over $C_a$.

As an ansatz for hypercharge  $G_4$-flux  we  consider the element 
\bea
G_4 = \sum x_{aI} [Z_{aI}]  \in H^{2,2}(\hat Y_4)
\eea
subject to (\ref{hyperG1}) and (\ref{hyperG2}).
In particular  (\ref{hyperG1}) implies that
\bea
\sum x_{aI} [C_a] v_{aI}[i] = 0  \quad {\rm for} \qquad i = 1,2,4,  \qquad  \sum x_{aI} [C_a] v_{aI}[i=3] =  f_Y \in H^{1,1}(S). 
\eea
This class $f_Y \in H^{1,1}(S)$ is furthermore subject to (\ref{hyperG2}).  Depending on the details of the fibration (\ref {transverse}) must be ensured as an extra constraint. 

Consider now a representation $R_a$ of $SU(5)$ and suppose it decomposes into $\oplus_k R_{a,k}$ with hypercharge $q_{a,k}$ under $SU(5) \rightarrow SU(3) \times U(1) \times U(1)_Y$. The matter surfaces ${\cal C}^{(n_k)}_{a}$ associated with these states are fibrations over $C_a$. 
The weight vectors associated to the representation $R_{a}$  split under GUT breaking into the weight vectors of the representations $R_{a,n}$. 
The hypercharge $q_{a, k}$ is then given by the inner product of $l_i$ introduced in (\ref{defwy}) with these weight vectors times a factor of $1/6$ from (\ref{defwy}).
Concretely, the charges obtained in this way are given in (\ref{decomp-rep}).
Then by construction the chiral index of the localised matter is given by
\bea \label{intqy}
\int_{{\cal C}^{(n_k)}_{a, k}} G_4 = \frac{6}{5} q_{a, n} \int_{C_a} f_Y,
\eea
where the result does not depend on the choice $n_k$ of course. The factor of $\frac{1}{5}$ is due to the factor of $5$ in (\ref{defwy}) which we have absorbed in $f_Y$.

Finally we turn to the quantisation condition and the twisting procedure.
For the $G_4$-flux to be properly quantised it must satisfy \cite{Witten:1996md}
\bea \label{FWcond}
G_4 + \frac{1}{2}c_2(\hat Y_4) \in H^4(\hat Y_4,\mathbb Z),
\eea 
which is tested by demanding that the integral of $G_4 + \frac{1}{2}c_2(\hat Y_4)$ over a basis of integral 4-cycles on $\hat Y_4$ be integer.
Let us ignore the possible half-integer shift in the quantization of $G_4$ due to the Freed-Witten anomaly if $\frac{1}{2}c_2(\hat Y_4)$ is not integer.
From (\ref{hyperG1}) $f_Y$ can take values in $H^{2}(S, {\mathbb Z})$, but this leads to fractional results from integration of $G_4$ over the matter surfaces in view of (\ref{intqy}) and the charges displayed in  (\ref{decomp-rep}). 
This can be remedied by a suitable fractional quantisation of the additional $G_4$-flux needed to produce non-trivial chirality. Since these fluxes do not break $SU(5)$ no constraints from   (\ref{hyperG1}) arise, and the only condition comes from integrality of the total flux integrated over the matter surfaces. 
This can be satisfied explicitly once the extra fluxes are specified e.g.\ as in \cite{Braun:2011zm,Marsano:2011hv,Krause:2011xj,Grimm:2011fx}.
Note that the requirement of integrality of $G_4$ integrated over the matter surfaces amounts to the criterion, put forward in IIB language \cite{Beasley:2008kw,Donagi:2008kj}, that the charged matter couples to integrally quantised gauge bundles, or more generally as a consequence of the Freed-Witten quantization condition as discussed in \cite{Blumenhagen:2008zz}. Indeed in $G_4$-language this condition is naturally seen to be the correct one in view of (\ref{FWcond}).

\section{Summary}

In this article we have studied the restriction of hypercharge flux to matter curves in intersecting brane models of type IIB string theory. The masslessness constraint on the hypercharge gauge field implies that its flux can only have a non-trivial restriction to components of the matter curves that are pullbacks of elements of the odd cohomology $H_-^{1,1}(X)$.  We have shown that this constraint, along with the D5- and D7-tadpoles, guarantees the cancellation of all anomalies with no other restrictions. In particular the cancellation of anomalies of the type ${\cal A}_{U(1)_Y-U(1)^2}$ proceeds through the orientifold-odd Green-Schwarz mechanism, which implies that the involved $U(1)$ gauge fields (other than $U(1)_Y$) are necessarily geometrically massive. We have exemplified this in a family of brane setups which admit non-trivial hypercharge restriction to the matter curves as required for flux-induced doublet-triplet splitting. To the extent that so far in all existing F-theory models a non-trivial restriction of hypercharge flux has only been \emph{postulated},  we have demonstrated for the first time that doublet-triplet  splitting by net hypercharge flux restriction can work consistently.
Our brane configurations are part of a whole class of such models in IIB which include and generalise the charge assignments found in F-theory $SU(5) \times U(1)$ compactifications. 

The calculations performed in this work are done in a type IIB string theory setting. However our primary aim is to deduce implications for F-theory models and in particular to study whether the anomaly ${\cal A}_{U(1)_Y-U(1)^2}$ must vanish in field theory. Our results show that this is not necessary if the $U(1)$s are geometrically massive in F-theory which, according to the studies \cite{Grimm:2010ez,Grimm:2011tb,Grimm:2013fua}, amounts to the statement that the $U(1)$ gauge potentials  arise by expanding $C_3$ with respect to non-closed forms on the Calabi-Yau four-fold associated with non-K{\"a}hler deformations. This result brings such $U(1)$s to the centre of model building in F-theory. It is important to note though that the primary interest for phenomenology is the flux along the $U(1)$s which induces chirality and, unlike the $U(1)$s themselves, can arise from harmonic four-forms. Indeed in \cite{Krause:2012yh} it was shown that for F-theory models with a IIB limit the so-called universal spectral cover flux of \cite{Marsano:2011hv} is precisely the flux along such a geometrically massive $U(1)$.

Another important result with implications for F-theory models is that anomalies of type ${\cal A}_{U(1)_Y^2-U(1)}$ can be induced by hypercharge flux. This implies that in F-theory models with geometrically massive $U(1)$s the constraints (\ref{dp}) can be relaxed. This opens up the possibility of evading the problem raised in \cite{Dudas:2010zb} regarding exotics and $U(1)_{PQ}$ symmetry. Indeed we constructed a toy example which does precisely this: it exhibits a massive $U(1)_{PQ}$ symmetry, which acts as a global symmetry at low energies and can only be broken non-perturbatively, while also having doublet-triplet splitting by hypercharge flux and no exotics.

Because the results are derived in a IIB setting they give only partial insights into their F-theory counterparts and leave many questions for future study. A practically important question is whether all the anomalies of type ${\cal A}_{U(1)_Y-U(1)^2}$ can be cancelled by this mechanism. In type IIB it is only the anomalies which mix the diagonal $U(1)$ of $U(5)=SU(5) \times U(1)$ with the other $U(1)$s that can be not vanishing, and this matches perfectly the fact that the trace structure of the operator relevant for the Green-Schwarz mechanism can only be non-vanishing for the diagonal $U(1)$ of $U(5)$. In local F-theory models however the type of anomalies that can be induced in the massless spectrum are not restricted to such a subset. Correspondingly, we would also expect that the trace structure can be modified, for example from Heterotic/F-theory duality one would expect that a trace over the full $E_8$ is possible. However the analogue of the geometrically massive $U(1)$s in heterotic compactifications is obscure and so the question of which anomalies are forced to vanish through geometric constraints, and so must do so at the massless spectrum level, and which can be cancelled by the GS mechanism remains open.

More generally the geometric mechanism for anomaly cancellation itself in F-theory, i.e.\ the uplift of the orientifold odd Green-Schwarz mechanism, remains as yet poorly understood. Given the important role we have highlighted for it, any understanding of its microscopics, perhaps through an approach along the lines of \cite{Cvetic:2012xn}, would be welcome. 

At a deeper level the fact that the constraint (\ref{dp}) was violated in explicit models shows that the uplift to F-theory of hypercharge flux restriction to orientifold odd components in IIB is not nescessarily related to the hypercharge restriction due to a local splitting, as discussed in section \ref{sec:hypf}. The F-theory realisation of this restriction will likely involve 3/5-chains or cycles as these are the natural uplifts of orientifold odd curves. How these combine with matter curves to allow for net hypercharge flux restriction is a very interesting topic for further study. It is possible that a whole new mechanism of hypercharge restriction in F-theory can be identified as the uplift of our IIB results.

Our understanding of the restriction of hypercharge flux to matter curves also has interesting implications for the flux-induced tree-level gauge non-universal corrections to the gauge kinetic function. We have shown that a net restriction to the matter curves must induce a moduli dependent splitting of the gauge coupling. Also, independently of this, we have argued that an alternative twisting procedure reduces the tree-level split of the gauge couplings observed in \cite{Blumenhagen:2008aw} by a factor of 5 to well within the 2-loop split within the MSSM. This is encouraging in the sense that we need not rely on particular cancellation coming from strong coupling effects for compatibility with precise tree-level unification. Clearly a proper M/F-theoretic computation of these coupling corrections, possibly similar to \cite{Donagi:2008kj,Dolan:2011aq,Grimm:2012rg}, would be desirable in order to settle more precisely the important question of gauge coupling unification in fully realistic F-theory GUT models.

Much remains  to understand in F-theory concerning anomaly cancellation and hypercharge flux restriction to matter curves. In this note we have made some progress towards this aim. Although modest,  it nonetheless has important implications for model building because the requirement of imposing anomaly cancellation for anomalies of the type ${\cal A}_{U(1)_Y-U(1)^2}$ is an extremely strong one \cite{Palti:2012dd} and left very few possible models. The possibility of using geometrically massive $U(1)$s to cancel such anomalies implies that many local models in the literature, for which the anomaly did not vanish, can be considered viable provided they have some global completion where the $U(1)$s are geometrically massive. Further, the problem of exotics in models with a $U(1)_{PQ}$ symmetry has plagued many constructions and we have shown how this can be avoided.

\subsection*{Acknowledgements}

We thank Mark Goodsell, Thomas Grimm, Sakura Sch{\"a}fer-Nameki and Martijn Wijnholt as well as Christian Pehle for very useful and enjoyable discussions.
The research of EP is supported by a Marie Curie Intra European Fellowship within the 7th European Community Framework Programme.
The work of CM and TW was funded in part by the DFG under Transregio TR 33 "The Dark Universe".

\appendix

\end{document}